# AN APPROACH FOR INTEGRATION TESTING IN ONLINE RETAIL APPLICATIONS

Roopa Singh[1] and Imran Akhtar Khan[2]

[1]Dept. of Computer Engineering and IT, JJT University, Jhunjhunu, Rajasthan
`roopas1983@gmail.com`
[2] Dept. of Computer Engineering and IT, JJT University, Jhunjhunu, Rajasthan
`imran4bc@gmail.com`

## ABSTRACT

*Retail applications has majorly fraud prevention, procurement, shipping and tax related, pricing, real time bank authentication applications integrated to make the application run successfully. Integration testing here plays an important role as it requires that all applications interact with each other and also interact correctly so that Retailer is at benefit. Different testing techniques and types are used to test the application. Different testing teams will perform integration testing but what is the correct approach and how should you proceed is the major concern of many. Here we propose at what stage of Software Testing Life Cycle (STLC), integration testing should be initiated. Also what should be the approach of performing the testing?*

*An example on Online Retail Application is used to understand the approach.*

## KEYWORDS

*Integration Testing (I&T), Retail, Testing, Software Testing Life Cycle (STLC), RTLOG (Retail Transaction Log), OMS (Order Management System), OLS (Online System), CC (Credit Card)*

## 1. INTRODUCTION

"The secret of successful retailing is to give your customers what they want. And really, if you think about it from the point of view of the customer, you want everything: a wide assortment of good quality merchandise; the lowest possible prices; guaranteed satisfaction with what you buy; friendly, knowledgeable service; convenient hours; free parking; a pleasant shopping experience."
- Sam Walton, World Famous Founder of Wal-Mart Corporation. [1]

With the fastest growth of Retail in 21st century, it is important that the Retailers provide the costumers with the most effective, user friendly and promising application. This is a challenging opportunity for them and for the team who performs the testing of such complicated applications. For testing team, it is always a concern what type of testing is to be performed to reach to/beyond the expectations of the customers. They face a challenge on – what are the different interacting applications, how they interact, what should be the response from to/fro application, is the response sent/received correct, and the challenge keeps on going. Additionally, the testing team has to know and understand the different testing techniques to come up with the plan which should help them to start and achieve the goal. This is a very tedious job and requires a lot more understanding and focussed approach.

In this paper, discussion is over the different techniques involved in the Web Based Retail Application. In a real world scenario, testing techniques and approaches vary from customer's application. Also each testing type is not important in all testing projects. So, it is important for a tester to understand the significance of each testing type and technique and their applicability. Here, our focus is on to understand the significance of Integration Testing in the Retail Web application. Also, the objective is to understand where I&T (Integration Testing) fits in the V-model. Finally the integration testing approach is being discussed using a real-time business case for the Retail Application

## 2. WHAT IS RETAIL?

**Retail involves the sale of goods from a single point (malls, markets, department stores etc) directly to the consumer in small quantities for his end use**. In a layman's language, retailing is nothing but transaction of goods between the seller and the end user as a single unit (piece) or in small quantities to satisfy the needs of the individual and for his direct consumption. Sometimes, the retailers purchase goods in bulk quantities (huge numbers) to be sold to the end-users either directly from the manufacturers or through a wholesaler [2]

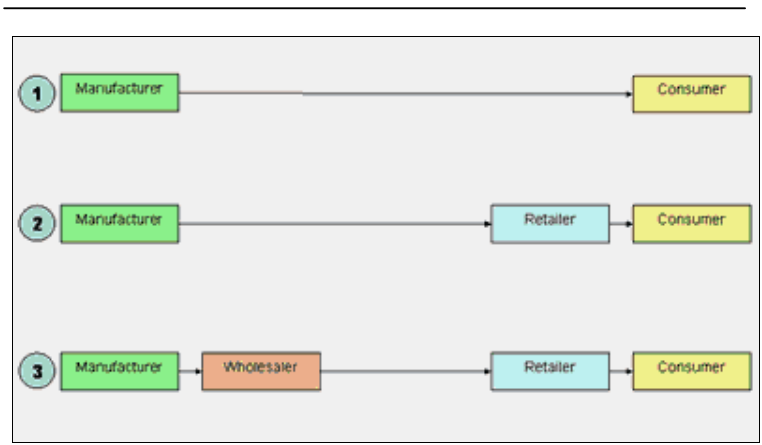

Figure1. Retail Channels

Online Retail Application (ORA) means transfer of goods from retailer/wholesaler to consumer by means of internet (web application of the retailer). The benefits (from customer's perspective) of ORA would be

- Wide variety of goods/merchandise with comparison
- Price comparison
- 24x7 shopping option
- Good deals and promotion

### 2.1. Need for Testing Retail Application

Retail industries face many challenges related to ongoing changing market trends, customer demand, and market globalization. When it's all about transformation and swift change, retail systems and applications become business critical and there's an increased need to provide competitive benefits. With little margin for error, retail industries need experienced partners who can perform unbiased testing and IT Quality Assurance.[4]

To ensure success retailers need to employ testing and quality assurance solutions that cover the IT landscape as a whole, including in-store solutions, enterprise management, eCommerce, and warehouse management. A success testing paradigm includes:

- Increase Application Quality
- Accelerate Time-To- Market
- Minimize Testing Cost

As per **Pareto principle** in software testing "80 percent of the uncovered errors in testing is from the 20 percent of the software components" (80:20 Rule) [3]. This implies that testing should be done thoroughly and in its initial stage to identify the 20% of the critical defects in order to avoid unexpected failure occurring due to 80% of defects. Hence, it is important to understand the different testing techniques and testing types to be implemented to test the Business-To-Customer (B2C) Web Application.

## 3. WEB APPLICATION: TESTING TECHNIQUES

A lot of effort and experience is required to understand the testing techniques to implement in the Web Application. In addition, testing technique for the integration testing is discussed. Integration testing technique approach is a mix of many. We cannot proceed with one in order to certify the application that it is well integrated. The various techniques applicable for I&T are

- Top Down Testing
- Bottom Up Testing
- Big Bang Testing

Here we'll understand the techniques and then with these, how the interacting application is sending the response. Is the response correct, is the response received and interpreted correct and so on. Let us consider each of them in detail

### 3.1. Top Down Testing

Top down Testing is an approach to integrated testing where the top integrated modules are tested and the branch of the module is tested step by step until the end of the related module. [6] This allows high-level logic and data flow to be tested early in the process and it tends to minimize the need for drivers. However, the need for stubs complicates test management and low-level utilities are tested relatively late in the development cycle. [16]

#### 3.1.1. Advantages

The advantages of top down testing can be

- Does not require drivers to be written
- Provides early working module of the program and so design defects can be found and corrected early [17]

### 3.1.2. Disadvantages

The disadvantages can be summarized as

- It enables testing only for the limited functionality
- Stubs are required in a way that it takes care of the integrated functionality
- Performed by Developers, which wastes time.

## 3.2. Bottom Up Testing

Bottom Up testing is an approach to integrated testing where the lowest level components are tested first, then used to facilitate the testing of higher level components. The process is repeated until the component at the top of the hierarchy is tested. [6]

### 3.2.1. Advantages

The advantages of the Bottom Up Testing are:

- The objects to be tested are known to the developer. So, it is easy to understand the scope of the test case creation and test data.
- Psychologically more satisfying because the tester can be certain that the foundations for the test objects have been tested in full detail. [18]
- Does not require stubs to be created

### 3.2.2. Disadvantages

The disadvantages are summarized below:

- The quality of the software can be guaranteed only when the testing is fully completed. Issue is, defects in the upper levels are detected very late
- Drivers creates complication in the test management
- Testing individual levels also inflicts high costs for providing a suitable test environment [18]

## 3.3. Big Bang Testing

Big Bang Testing is an approach to Integration Testing where all or most of the units are combined together and tested at one go. This approach is taken when the testing team receives the entire software in a bundle. So what is the difference between Big Bang Integration Testing and System Testing? Well, the former tests only the interactions between the units while the latter tests the entire system. [19]

This is the approach used by the QA Integration Testing team.

### 3.3.1. Advantages

The advantages of the Big Bang Testing are described below

- Most of the integration related defects are identified in real time, proper integrated environment

- Does not require effort by development for creating the stubs and drivers
- Provides better coverage and efficient testing than traditional bottom-up and top-down testing
- Helps to provide the customer with better integrated environment

### 3.3.2. Disadvantages

The disadvantages of the Big Bang Testing can be summarized as:

- It is harder to track down the causes for the errors since all the modules, and thus complexity, are added at once. For instance, if a transaction is not processing correctly, you may have to track back through 10 modules to determine the cause. [20]
- Integration testing cannot be started until all the modules are completely created and tested
- It is difficult to identify where in the module the error is.

We have discussed the techniques which are used usually but QA team uses the Big Bang. But the most important technique can be used is the experience with error guessing. We'll discuss the same below

### 3.4. Error Guessing

Error Guessing does not depend on any rule which could be applied. This is all about your experience in the domain. The more experience you have, the more defect areas you can analyse which could not predicted by the tester who is brand new to the application. Integration testing in the web application needs more of the experience to identify the test conditions which could break the system.

Techniques that include error guessing would include the following

- Knowledge of the application under test (AUT)
- Knowledge of the related testing types done (to understand where defects were found during the test)
- Knowledge of the integrating applications with their behaviour of response

### 3.4.1. Example of Error Guessing

Let us consider the Online Retail Application which integrates with the Taxing system and the bank system in real time. So based on the experience you may consider the following condition to test:

- Place the order when the taxation system is down. Test what is happening to order?
- Try to place the order with card which is going to be expired?
- Try to place the order when the bank is UP but the web service is down. How the order would be managed by the system?
- Try to click on the back in the browser when performing the payment of the order

There could be so many conditions which could be guessed and also the system breaks. The more the integrating applications, the more defects prone the application is.

## 4. APPROACH TO PERFORM INTEGRATION TEST IN WEB APPLICATION

Testing is a very challenging task and needs a lot of planning before you start the testing. Also while testing there are numerous things to be taken care of in order to identify the defects, which improves the quality of the system.

"Testing is the process of exercising or evaluating a system or system component by manual or automated means to verify that it satisfies specified requirements, or to identify differences between expected and actual results." **IEEE**

Testing methodology that we have considered is V&V Model. In this model, testing team works in parallel with the development team. This provides the testing team sufficient time for preparation in order to deliver the application with high quality.

**Validation:** Ensures the building of right product

**Verification:** Ensures that the product is right

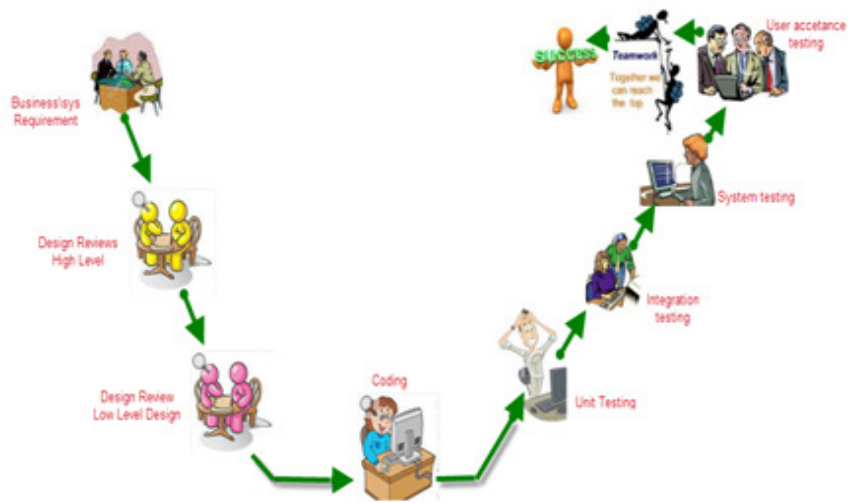

Figure 2. V&V Model

So where does the Integration testing fit? This is an important question to be answered. To make the integration testing more effective, it is good that QA team start once 70-80% of the system testing is completed with major areas passed. Unless this happens, it is of no use that integration team starts their testing. So, Integration testing would be a step before UAT (User Acceptance Testing) or Production testing. It is this team which certifies the product quality.

### 4.1. Requirement Analysis

Requirement Analysis which is the first phase of the Testing Life Cycle is very critical for a tester, which will indeed help you in planning and construction phase. Things which are to be taken care are

International Journal of Computer Science & Information Technology (IJCSIT) Vol 4, No 3, June 2012

- Identify and understand what the requirement is?
- What are the integrating applications?
- Are the requirements feasible to test?
- How much time and effort will be required?
- Identify the gap in requirements, if any.
- What are the limitations?
- What are the risks in terms of testing?

**4.2. Test Requirements**

Identifying the requirements correctly for the integration testing is a challenging part. All the requirements of the project for a system would not be eligible for I&T. Out of the requirements which have a mix of system and integration, identify the requirement/(s) which may cover the end to end flow in the applications. Once you have identified, testing team looks for the various documents which will help them to understand the requirements better.

Below are the documents to be referred:

- **Requirement Specification Document**

    This could be Market Business document (MRD), Business Requirement Document (BRD) or a combination of both (MRD + BRD)
- **Functional Requirement Document**

    This document will describe how the requirements will be working ie the function in the system after the implementation of the requirements. This may also have the Use Cases to be used by the team
- **Design Specification Document**

    This document will help you in identifying the design changes. What are the table details, Web service request and response, how is the flow of information between the integrated applications, when the different applications will be integrating and where the data/information will flow.

After going through the above documents, it becomes easier for the tester to identify the valid integration test scenarios and conditions.

**4.3. Test Planning**

"Failing to plan is a plan to Fail" **Effie Jones [8]**

During this phase of the STLC, following questions will be answered

- **What will be tested and what will not be tested?**

    Out of the n-applications, there could be some applications which are not accessible (may be due to the fact that the application is owned by third-party). These applications may not be available for testing. So, identify the applications and functionalities which could be tested for your integration.

147

Eg: If you want to test the scenario when the interacting application say BANK is down, then you could not test this as it is real time and you cannot make the application down as it has impact on the Bank.

- **What would be the strategy to test the application?**

A **test strategy** is an outline that describes the testing approach of the software development cycle. It is created to inform project managers, testers, and developers about some key issues of the testing process. This includes the testing objective, methods of testing new functions, total time and resources required for the project, and the testing environment.[9]

This would include how to proceed with the testing. Are we planning for the smoke test with end-to-end before we actually start the actual integration testing? How are we going to proceed if plan A fails?

- **How many resources will be required to test?**

Based on the project requirements and the complexity, test manager needs to identify that how many resources will be needed to work on the integration testing phase.

- **Role and responsibilities of each member in the team**

Team Lead or Team Manager will need to identify the Point of Contact (POC) for the speciality skills on the application to be tested. Before starting the execution or going into the design phase, it should be clear to each team member what the responsibility is in each phase and how they should proceed.

- **What would be the environment?** [7]

It is important to know what the environment to be used by the team is. Not all the applications would be integrated to all the applications. So to proceed further, it should be well integrated.

- **How many cycles will be performed to complete the testing?**

Based on the end date of the testing and the time when you are starting the integration testing, it is important to identify how many cycles will be performed to complete the cycle. Each cycle will have the entry and exit criteria. In real IT testing phase, 3 cycles are the standard to certify that at least 90% test cases have passed.

- **Entry and exit criteria for the Integration testing?**

Usually it is a myth that integration is before the system testing. But inorder to test the integration of the application, I&T can start only when

> Almost 80-90% of the system testing is completed and the related test cases have passed.
> Smoke test for end-to-end test case is passed
> All applications are integrated

**When to stop I&T?**

- All the cycles (usually 3) are completed
- Entry and exit of each cycle is met
- Regression on the integrated applications is performed when the code is freezed.

- **Timelines for the integration testing to start and end**

When the Integration is good to start and when it should stop should be decided upfront. If that does not happen, it would mess up the whole deployment process and would end up with date extension to the production environment.

- **What tool will be used to log the defect and what would be role of each team while logging defect?**

It is to be finalized that what defect management tool will be used (HPQC, ClearQuest). Each team member has different role to play. Restricted right is given to change the defect status so that things does not get messed up. Also, the process of how to log the defect will be identified in order to make the process smooth and common across the team. This will reduce the confusion within the team too.

- **Identification of the risk, mitigation plan**

Team Manager and Team Lead work together to understand the risk the project can have with the limitations on time or application access. This has to be informed to the higher management before starting the integration testing execution. Also, the mitigation plan has to be proposed to the management so that they know how we are going to proceed in case Plan A does not work out. We may have to go with the Plan B which could have more risk.

All the above, will result into two important deliverables which are provided the Team Manager or Team Lead. The deliverables are

- Test Plan
- Test Strategy

### 4.4. Test Design/Test Construction

Test Design starts with the identification of the test scenarios from the integration testing requirements. It can be diagrammatically represented as

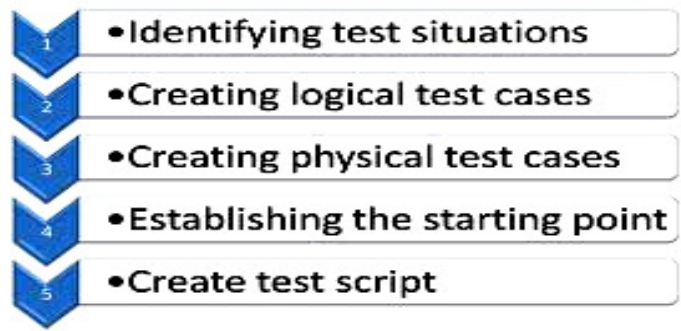

Figure3. Steps for Test Design

"Good test design leads to Successful tests. A test that reveals a problem is a SUCCESS" [10] Test Design is the very critical phase for the tester/test team. This phase includes the Test Validation activities. The various activities included are:

- **Requirement Traceability Matrix (RTM) and Test Coverage [7]**

    RTM is the document which helps to trace the test deliverables. It can be used for the following

➢ Track all requirements and verify that whether those are covered in the design phase[11]
➢ Helps to create the test deliverables including the test script
➢ In case some requirement changes, it is easy to trace back and trace forward and update the test scenario/condition/case accordingly. This also ensures that none of the requirements are uncovered for the testing

- **Test Scenario and Test Condition Identification**

This involves a lot more deep knowledge of the integration end-to-end flow. Test Scenarios are the deliverables for the business process flow that are to be tested. There could be one to many relationships with the test scenario and the conditions which could be even positive and negative.

With respect to integration testing, it is important that tester is able to identify the impacted system and the high level flow. Also, these deliverables could be related to one requirement which is impacting many applications or many requirements which could be combined to create one scenario or condition. Also, consider the offline scenarios which are important to be covered in the web based applications. This is the point where many applications break.

- **Test Case Preparation** [7]

    Once the test condition is identified and the expectation out of it is provided, it is important to understand the condition correctly so that you proceed with all validation steps which are important part of your condition.

    Thing to remember while creating test case for I&T phase:

➢ Check for the request and response of each interacting application
➢ Validate the same case with large amount of test data and also different data
➢ Check for the value what the integrating application/interfacing application should have versus what it is currently having. Is that correct? Sometimes the interfacing applications interpret the request in different way and respond wrong but do not fail. Be careful!
➢ In case you have database validation, check for the columns which are affecting with the different data. Don't ignore any of the tables which are impacting the transaction/flow
➢ Check for the error logs. Sometimes there are error logged even though the transaction is flown correctly in the interfacing application
➢ Provide the proper description of the test case with the changes in the interfacing application, this helps to understand the flow at a higher level
➢ If different data respond differently for the same test condition, provide that and don't forget to test that. Think Out of the Box to break the system
➢ Last but not least, when data flows from one application to another test immediately when you are expecting the data to flow. Don't think that it could be tested later. The more active you are with the data flow validation, more defects you have in your bucket.

- **Identification of Test Data**

    Managing test data in the non production environment is essential since it helps ensure proper validation of business process.[13] If you don't have any systematic approach for building the test data while writing and executing the test case, then there are chances of missing some bugs which comes when you have valid/invalid data as per the condition.[12] So, it is suggested that you identify the test data while you are identifying the test condition. In case that is missed out, don't miss while you are writing the test case. Be specific, and destructive.

- **Test Case Review and Approval** [7]

    Test cases are one of the most widely used test artefacts in software testing activity. More often than not they decide the effectiveness of the testing efforts. With test cases playing such an important role, it becomes necessary to ensure that test cases are effective, precise, and optimized. This is when a test case review process helps.[14] Review process can be diagrammatically represented as

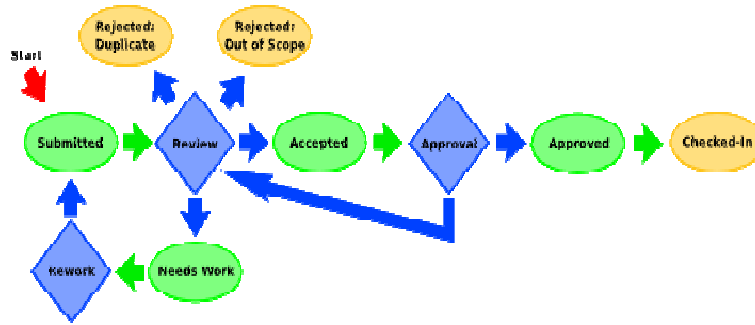

Figure4. Review Process

With respect to I&T, reviewer should consider the following

- Are the test cases integration related?
- Can the test cases be merged with some other test case (if any)
- Are the test steps written in a way that even a new person can execute or needs more detail?
- Check for the scope of I&T
- Check for the flow of information mentioned in the test case. It should properly sequence with proper validation points. If this is missing, no point in testing the integration part.
- Also, get the test cases reviewed by the functional expert who can better understand the scope and flow.
- Get the review comments jotted down so that it is made sure that you can track that none of the review comment is not being incorporated
- Standardise the test case format and follow the same format across the team. It helps to understand the flow when you are reviewing the test case written by any one.
- Check for the correct severity and priority of the test case. This helps when you have less time and need to identify the test cases which could uncover the more defects with less test cases (usually by severity basis). It is important for tester to understand the severity and priority categorization and take up the decision diligently

## 4.5. Test Execution

Testers execute the software based on the plans and test documents then report any errors found to the development team.[15] Following considerations should be done while performing the test execution:

- Execute the cases with correct data
- Don't skip the validation point. Try to add additional validation for the request/response/database check.
- Keep a track of the results for each test case in each life cycle.
- In case, difference between the actual and expected exists then move to the below step to raise the defect

## 4.6. Defect Tracking

When the difference between the expected and actual is observed, then tester is supposed to raise the defect. The defect cycle should be understood with the below diagram

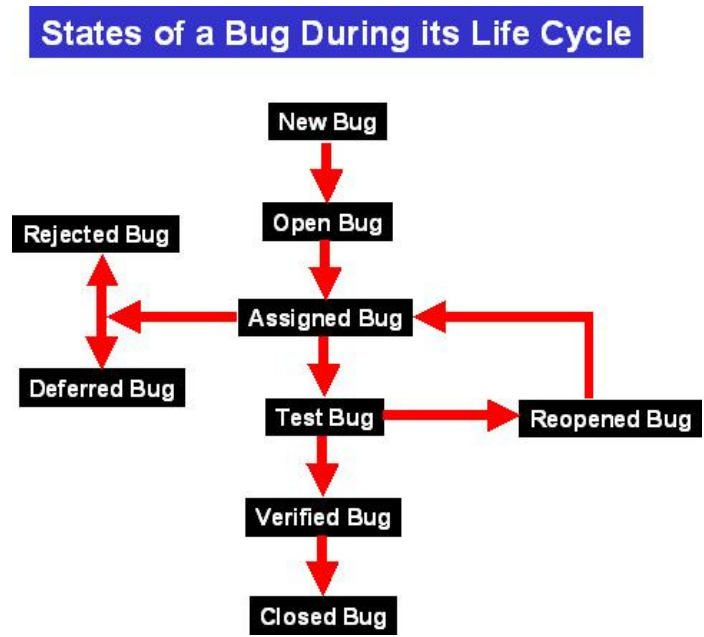

Figure 5. Defect Life Cycle

To raise the defects, below points should be taken care:

➢ Check if the defect is reproducible
➢ Provide the test description, steps to reproduce, actual and expected result correctly
➢ Provide the correct functional area and application
➢ Provide the environment details
➢ Provide the correct severity to the defect. Since you are performing I&T, the integration defects should be S1 or S2. Categorization should be done with proper care. If there are more than 15 test cases impacted with the defect (may be blocking the functionality) then it could

    be S1 rather than S2. So, it is very important to give correct severity based on which developers give priority to fix the defect.
- Try to give the request and response or test result
- Provide the correct test data used
- In case it is possible to give recording, provide it. This will help the developers understand the steps as each developer does not know the end-to-end functionality and how to reproduce the defect.
- Make sure the description with the application is quite clear and precise.
  Eg: [Env][Application] – Description of the defect

## 5. EXAMPLE FOR I&T OF WEB BASED APPLICATION

When we place an order in Online Retail application, it is not only the online system that is involved, but also Fraud Preventive, Payment System, Web Services, Order Management System (OMS) are involved which work at back end/front end to process the order.

Let us consider an end-to end flow of order placement. With this example, we'll understand how we should perform the integration testing of the interacting applications.

Below is the flow we'll be considering

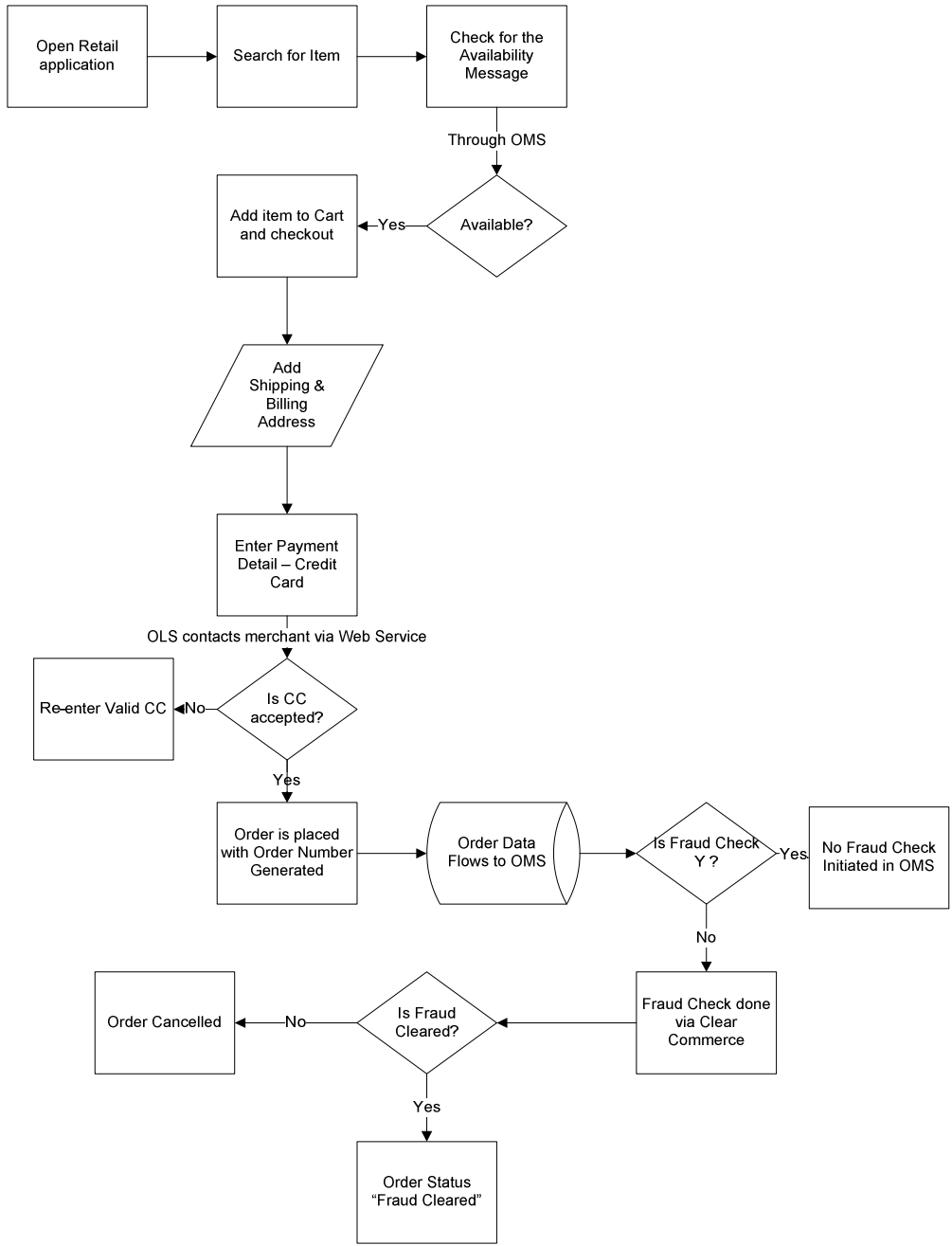

Figure6. Order Creation and Processing

## 5.1. Test Condition

Place order using credit card which does not lie in the BIN range of the VISA Credit Card when the fraud system is OFFLINE and Bank system is UP.

**Note:** A rule has been set in the fraud system such that customer IP address has been categorised in the fraud list.

### 5.2. Test Preparation

Before you start with the integration testing, below are the key points to be taken care of

   I. Identify the applications which all are integrating the test scenario ( Banking system, Web Service, Fraud preventive system, Order Management System, ReSA)
  II. How is the information flowing to and fro?
   - Identify what should be the data flowing when you add credit card which is not within the BIN range or valid credit card
   - What should be the response of the merchant/bank in each case
   - How the merchant and the online application is interacting (Web Service)
   - Where does the order flow after being placed in Online system
   - What is the value sent to Order Management system to identify if the fraud check is done or not
   - What should be the order status when the Order is in Fraud Negative list
 III. Identify the test data
  IV. Identify the applications you need to have access to.
   V. Check whether you have integrated environment

### 5.3. Test Case Creation

Below with the example of the test case we'll learn how the approach should be when we perform I&T.

**Objective:** The objective of the test case is to place order using credit card which does not lie in the BIN range of the VISA Credit Card when the fraud system is OFFLINE and Bank system is UP.

**Applications Involved:** Online system (OLS), Fraud Preventive tool (Clear Commerce), Merchant (Visa), Order Management System (OMS), ReSA (Retail Sales Audit)

**Pre-requisite:**

- Clear Commerce is OFFLINE

- Online System/OMS is UP and Running

- Merchant/Web Service is UP

- Total Stock on Hand of item is noted

Table1. Integration Test Case

| Step No. | Description | Expected | Application |
|---|---|---|---|
| 1 | Open the Online System and browse the item | Item is available in the Online System | OLS |
| 2 | Check for the Inventory | Availability message is displayed as "Available" | OLS > OMS |
| 3 | Add item to the Cart and checkout | Added item is checkout | OLS |
| 4 | Add Shipping and Billing address | Shipping and billing address are added | OLS |
| 5 | Add credit card details in Payment Information page and click on continue<br>- Credit card type: VISA<br>- Credit card: 7978998767854345 (which is not within the VISA range<br>- Expiry: 05/2012 | | |
| 5.1 | Validate that Credit Card details are masked | Masked CC details are sent to Merchant via Web Service | OLS > Merchant via Web Service |
| 5.2 | Validate that Credit Card details are correct as that entered | CC details are correct | OLS > Merchant via Web Service |
| 5.3 | Merchant validates the Credit Card and send the proper decline message | Response Code "227" returned with reason "Merchant cannot accept this Private Label BIN range" | Merchant > Web Service |
| 5.4 | Validate the error message in the Online system | Proper error message is displayed | Web Service > OLS |
| 6 | Change credit card number in Payment Information page and click on continue<br>- Credit card: 4213238767854345 (which is within the VISA range<br>- Expiry: 05/2012 | | |
| 61 | Validate that Credit Card details are masked | Masked CC details are sent to Merchant via Web Service | OLS > Merchant via Web Service |
| 6.2 | Validate that Credit Card details are correct as that entered | CC details are correct | OLS > Merchant via Web Service |
| 6.3 | Merchant validates the Credit Card and send the acceptance message | Response Code "00" returned with proper AVS reason description | Merchant > Web Service |
| 7 | Place the Order | Order number is generated | OLS |
| 8 | Validate the XML that provides information to OMS | Order Number, Order Details, Fraud Check indicator "N", Order Status " Created" | OLS > OMS |
| 9 | Validate the response when OMS contacts Clear Commerce | Fraud Cleared "N" is set with the rule getting triggered | OMS > Clear Commerce |
| 10 | Validate the status of order in OMS and OLS | Order Status "Cancelled" in both system | OMS > OLS |
| 11 | Verify the RTLOG generated | RTLOG has the status as ORDC (Order Cancelled)  for the order | OMS > RTLOG |
| 12 | Upload the RTLOG in ReSA | No inventory change for the item | ReSA |

With the test steps mentioned above, we have considered the validation for each of the processing happening within different interacting application. It ensures that when any of your action is interacting the other system, what is the request sent, what is the service used to interact and what is the response. This is where you identify the defects. Usually the request and response comes different, which is more common in the online applications.

## 6. CONCLUSIONS

Integration testing is most important phase of the STLC which is also difficult to perform. Performing integration testing of all the related applications is quite different from just testing the flow within just two interacting applications. In real time, testing team perform this without the stubs or drivers. The integrated environment is required for the integration QA team to start with the testing. With this paper, it helps us understand what the importance of the integration testing is and what should the integration testing team do to give better results to their effort.

Retail web applications are becoming the most preferred way to do shopping. So, to provide customer with satisfaction and err free application, it is quite important to perform integration testing as this will remove the bad experience from customers. QA team will identify that and development team will fix it. This is the reason; we also understood what should be the approach when performing I&T in Retail Online application. This will help the QA team to understand how the validations should be done and what should be the consideration in each of the scenario. Understanding the approach is an important part. There are still areas where we can improve based off the experience as experience is the best master of all.

## Authors

**Roopa Singh** is currently a research scholar in Computer Science from JJT University, Jhunjhunu, Rajasthan. Her area of interest is – E-Commerce (B2C and B2B), Web based: System and Integration testing. She has received her Master of Computer Science and Applications (MCA) degree from Aligarh Muslim University, Aligarh.

**Imran Akhtar Khan**, is pursuing PhD in Computer Science from JJT University, Jhunjhunu, Rajasthan. His research is Software Testing- especially skilled in Integration testing, Web based testing. He has completed his Master of Computer Science and Applications (MCA) degree from Aligarh Muslim University (AMU), Aligarh.